\documentclass{nime-alternate}

\usepackage[utf8]{inputenc}
\usepackage{multirow}
\usepackage{subfigure}
\usepackage{hyperref}

\begin{document}
%
%
%\conferenceinfo{NIME'14,}{June 30 -- July 03, 2014, Goldsmiths, University of London, UK.}
\conferenceinfo{}{}
\title{Designing Sound Collaboratively -\\ Perceptually Motivated Audio Synthesis}

\numberofauthors{3} 
\author{
\alignauthor Niklas Kl\"ugel\\
      \affaddr{Technische Universität München}\\
       \affaddr{Department of Informatics}\\
       \affaddr{Boltzmannstrasse 3}\\
       \affaddr{85748 Garching bei München}\\
       \email{kluegel@in.tum.de}
\alignauthor Timo Becker\\
      \affaddr{\mbox{Ludwig-Maximilians-Universität} München}\\
       \affaddr{Department of Informatics}\\
       \affaddr{Amalienstrasse 17}\\
       \affaddr{80333 München}\\
       \email{timo.becker@campus.lmu.de}
\alignauthor Georg Groh\\
          \affaddr{Technische Universität München}\\
       \affaddr{Department of Informatics}\\
       \affaddr{Boltzmannstrasse 3}\\
       \affaddr{85748 Garching bei München}\\
       \email{grohg@in.tum.de}
}

\maketitle
\begin{abstract}

In this contribution, we will discuss a prototype that allows a group of users to design sound collaboratively in real time using a multi-touch 
tabletop. We make use of a machine learning method to generate a mapping from perceptual audio features to synthesis parameters.
This mapping is then used for visualization and interaction. Finally, we discuss the results of a comparative evaluation study.
\end{abstract}
%\vspace{-0.2cm}
\keywords{\vspace{-0.05cm}Collaborative Music Making, Creativity Support, Machine Learning, Sound Design}
%
%
%\vspace{-0.2cm}
\section{Introduction}
%\vspace{-0.05cm}
Sound design is generally seen as the process of specifying, generating or manipulating
audio elements in a non-compositional context. If we look at it more from the musical perspective - sound design as designing timbres - then sound design is part of the daily routine in 
modern electronic music composition and performance as it is a form-bearing dimension of music \cite{Holmes2011}. 

Motivators for approaching such creative tasks can be endogeneous or exogeneous. 
The former encompasses the goal oriented creation of sound forms with functional intent. E.g.~from a compositional perspective this includes communicating the compositional structure. From a social and performative perspective, timbre forms a significant part of the musician's assessment of \textit{musical identity} since it is the primary medium of interaction with the other musicians \cite{Cottrell2004, Holmes2011}. Exogeneous motivators correspond to explorative creation of sound forms, "playing with sound for its own sake" without any initial presumptions. An outcome can be the extension of one's own musical vocabulary.

Technically, there is a plethora of methods to generate various timbres, such as Abstract-, Physical- and Spectral Synthesis or via processing recordings. For an in-depth review consult \cite{Zolzer2003}. In general, the parameters for the synthesis method are not grounded in the \textit{perceptual} domain, but rather in the structural one.
Hence, they may frequently exhibit non-linear behavior and inter-dependencies regarding the perceptual qualities of the output. 
Changing timbre is therefore non-intuitive without prior knowledge about this structure and its technical functioning \cite{Risset2003,Robel}. This is
especially true for novices. In a collaborative setting, this may be even more problematic as tasks involving group creativity greatly benefit from heterogeneous groups and therefore from
domain specific novices \cite{Fischer2006,Morgan1980,Uzzi2005} \cite[p. 137-144, p. 450-452]{Sternberg1999}. Hence, to foster social creativity, this heterogeneity must be integrated
\cite{Fischer2005}. This beneficial effect is further amplified by the phenomenon of \textit{Group Flow} \cite[p. 158]{Sawyer2006}. The experience of Flow itself can be defined as ``a
holistic sensation that people feel when they act with total involvement'' \cite{Csikszentmihalyi1990}. It is insofar important as it stimulates an implicit learning process, 
enables empathical involvement with the music \cite{Nijs2009} and is also bound to the feeling of \textit{social presence}. 
Group Flow as a social experience has been shown to foster objectively more valuable musical results \cite{MacDonald2006} and is a key success factor in Computer Supported
Collaborative Music making (CSCM) \cite{Swift2004}. In this way Group Flow is a motivator and means for the group to innovate in a creative task.  The engagement caused by 
intrinsic motivation also has the effect of supporting the learning
process for musical expression, thus contributing to mediating the interaction with the shared
CSCM environment and, finally, the social interaction with peer members.
Concludingly, we regard the integration of these social effects as highly beneficial for the creative task of creating music.
The utilization of a multi-touch table especially offers means to facilitate
social communication protocols \cite{Wigdor2011,Hornecker2007} and we will therefore use such a device in this contribution.

Certain key aspects have been identified for realizing collaborative
applications. Based on the work by Cockburn \cite{Cockburn:1995} and Dourish \cite{Dourish:1992}, the most prevalent ones within the scope of this contribution are:
\textit{Group Awareness}, which describes the ability of individuals in the group to share and gain mutual knowledge about the activities performed and
\textit{Group Articulation}, which refers to the ability to partition the group task into units that can be combined and substituted. %
Fostering these aspects will serve as basic requirement for our interaction design. 

It is further necessary to define the notion of \textit{timbre} that we will use. While timbre, in general, attributes certain perceptual qualities
to sound (e.g.\ brilliance), we use it here as an abstract descriptor of perceptually salient acoustic features that can be modelled and measured by IT systems.
In this regard, timbre can be a set of low-level features such as frequency components evolving over time or high-level features such as the spectral centroid.

The structure of this contribution is further divided into three parts, essentially: after discussing the relations of our contribution with the body of related work, we will introduce
the main elements of our approach. We will then elaborate on a comparative user study that we conducted to evaluate our approach and our prototypical implementation.
We will conclude with a discussion of the results and future work.
%
%
%
%
%
%
%\vspace{-0.2cm}
\section{Related work}
%\vspace{-0.05cm}
The majority of research related to mapping from timbre to synthesis parameters focuses on re- or cross-synthesizing the original sound sources from timbral features.
For our use-case, however, we are interested in applying this mapping foremost as a paradigm for interaction, so the main issue is human cognitive manageability. This field of Human Computer Interaction has not been studied extensively
\cite{seago2013new}.
Regarding the cross-synthesis methods, we can identify three common approaches, the signal processing one, which formulates explicit timbral transformations (e.g.\ \cite{Serra1997}), the machine learning one, where the timbral model 
is inferred from audio data (e.g. \cite{LeGroux2008}) and the concatenative one, where timbre is constructed using the sound recordings themselves.
A well known representative of the latter is the CataRT system \cite{Schwarz2012}. It is especially relevant as the navigation in timbre space
is its core concept for interaction. However, for this, CataRT and subsequent developments use the high-level timbral features as orthogonal axes directly. 
This means that only a low (2-3) number of features span this space, discarding timbral qualities that may add valuable information
and possibly reducing the representative quality of the space since features may be correlated. 
To circumvent these problems, CataRT allows the user to re-define the space at run-time. 
For the collaborative use-case, this introduces the conceptual challenge how a shared navigational space can have
user-specific views without hampering awareness and articulation. 
In case of the collaborative application Nuvolet \cite{Comajuncosas2011}, which is based on CataRT, 
this space is static. It uses a 3D camera system as sensor for gesture driven Audio Mosaicing.  Nuvolet aims to support a collaborative, virtuosic performance that enhances the performer-to-audience relationship.
Consequently, it lacks features that provide affordances for the collaboration itself with respect to shareability and awareness. The cognitive load
and imprecision that are initially induced by the 3D control method are also problematic for novices.

Apart from high-level timbral features, also low-level features that are reduced in dimensionality can be used to span the timbre space \cite{Nicol2004, LeGroux2008}. 
However, the constructed timbre space by this machine learning approach may not have an obvious relationship
to the human perception \cite{LeGroux2008}, so in order to remedy this, further processing steps than the reduction itself are necessary. Moreover, any mapping to synthesis parameters \textit{and} the inter-relationship between timbral features are bound to be highly non-linear \cite{Hoffman}. 
Non-linear mappings such as Self-Organizing Maps (SOMs) are therefore preferable. These have already been applied in numerous fields including %
sound-design. %
Especially the approach in \cite{Eigenfeldt} overlaps with our use-case as the generated 2D-SOM representation of the timbre-space
is used as interaction metaphor. Due to some beneficial mathematical properties, we will use Generative Topographic Mapping (GTM) instead of SOM, as will be explained in greater
detail in section \ref{mv}. Furthermore, we will use high-level features since otherwise we would have the additional problem of defining a proper mapping that is computationally expressive enough to abstract from low-level features to perceptually more meaningful ones. 
To conclude, we see the collaborative use-case of designing sound a largely untapped territory for fostering creative endeavors.

%
%
%
%
%
%
%
%
%
%
%\vspace{-0.2cm}
\section{Synthesis \& Corpus}\label{sc}
%\vspace{-0.05cm}
We need to be able to synthesize a large variety of different timbres %
in order to not severely limit expressivity (apart from the constraints imposed by the sound generating method). It is the goal to generate a corpus $\mathcal{S}$ such
that each sound $s_i \in \mathcal{S}$ can be analyzed for its timbral features $t_i \in \mathcal{T}$ to generate the mapping $s_i \leftrightarrow t_i$.
Later on, we would like to synthesize $s_i$ again in real time within a tonal context. %
A common approach would be to use and re-synthesize recorded sounds. However, it is non-trivial to analyze all time-frequency relationships reliably for an
\emph{arbitrary} sound source, therefore possibly rendering its original timbre-space relationship invalid during re-synthesis (e.g.\ when the fundamental
pitch is altered). Therefore we opted to develop our own synthesis model such that all parameters of the synthesis (e.g.\ pitch) are known.

To create the corpus it is therefore necessary to sample various parameter settings $p_i \in \mathcal{P}$ from the possible parameter
combinations $\mathcal{P}$ to generate the sounds $s_i$ for the analysis. Hence, we are interested in having a low number of parameters. We used Vector
Phase Shaping (VPS) \cite{Kleimola2011} at the core of our synthesis model. It is an abstract synthesis method employing phase distortion and allows for the generation of a variety of harmonically rich waveforms as well as filter phenomena (e.g.\ formants) using only two parameters. This is achieved by modifying the phase function using a singular two dimensional break-point that may be moved beyond unit phase.
We used two VPS based oscillators in a master-slave configuration and a noise source. The slave oscillator can be set in harmonic intervals relative to the master oscillator's pitch. Both can be either mixed or set to amplitude or frequency modulation. 
The sounds generated can evolve over time as most of the involved synthesis parameters (e.g.\ VCA and FM index) may be modulated using triggered envelopes. %
Furthermore, to blur or emphasize  spectral peaks we added a flexible effects chain (reverb, chorus and flanger). With this configuration the synthesizer is able
to create a variety of timbres such as bassy, percussive, leading and atmospheric ones. 
Further compacting 
the parameterization led to $16$ parameters for the synthesizer that were then individually discretized to at most $20$ steps. This resulted in $|\mathcal{P}| \approx 10^{15}$ 
possible parameter combinations.

Ideally, we would like to cover the whole extent of timbral varieties that can be achieved with the proposed synthesizer for the corpus.
However, given the cardinality of $\mathcal{P}$, it becomes clear that synthesizing all parameter configurations is not feasible.
Hence, the corpus can be created either by an expert, automatically by performing some high dimensional search method or a mixture of both. We applied the latter, 
as a heuristic method which makes use of audio similarity. $\mathcal{P}$ forms a hypercube if we interpret the normalized parameters as orthogonal axes.
Thus, within the volume of such a hypercube the set of points represent valid parameter configurations. 
A hypercube can be further split along any axis, creating two siblings with a smaller volume and less included parameter configurations. 
Furthermore, we can determine an estimate of the similarity of the sounds within a hypercube if we measure the similarity for the \textit{synthesized} audio that two parameter configurations at opposing vertices on each side of such a split represent.
Using this heuristic, we are essentially performing a multi-dimensional binary search, creating dissimilar sounds along a path that eventually leads to a recursively generated volume  containing mostly similar ones. We can gain a speed-up for the search method as the produced siblings are independent and can be processed in parallel. 
Given a set of parameter configurations - or presets - that an expert has created, the method can search for sounds in between these according to the similarity
measure; each pair of presets is then interpreted as two extreme vertices describing a hypercube uniquely. We used the similarity measure proposed in
\cite{Pampalk2004}, as its robustness to arbitrary sounds has been shown. The nature of VPS to predictably generate, for our ears, mostly musical spectral effects 
makes this method applicable. Note that the fundamental pitch for each sound has been normalized to C, differing only in the octave (-5, 5) for the whole data set.
The length of the generated samples for the subsequent feature analysis has been set to 4 seconds.

%
%
%
%
%\vspace{-0.2cm}
\section{Features}
%\vspace{-0.05cm}
%
%
%
We performed the analysis of the corpus to derive only high-level timbral features exactly the same way as described in \cite{Klugel2013} using the MIRToolbox \cite{Lartillot2008}.
These generated features are largely time series data, which, in our use-case, are difficult to integrate: First technically difficult, as the synthesis uses modulators whose state would have to be saved with the generated audio in order to properly represent every frame of a time-series feature in a $t_i \leftrightarrow p_i$ mapping. Second, conceptually difficult, as
each sound in the timbre space is then represented as a path. A visualization of this may overburden users, but it also requires a more complex user interface and more interaction as path operations are now the canonical way of exploring and designing sound. And third, practically difficult, as the memory requirements for the visualization method is $\mathcal O(n^2)$ and our
feature data set exceeds $100$GB.
Thus we collapsed the originally 30 dimensional time series data onto a single 368 dimensional feature vector by extracting statistical properties of the features.
Since the projection method, which will be introduced shortly, is not well suited for this high dimensionality of input data, we performed a greedy forward feature selection yielding 50 features. Furthermore, the original synthesis parameters were added to the feature vector in order to include to some degree 
information about the temporal evolution of a sound. Initial experiments by listening tests using the initial prototypical application revealed that this has a significantly positive impact on the quality of the mapping. We allot this finding to reducing the amount of ambiguous information for the entire feature vector, as the conversion reduced the temporal
descriptiveness for some of the original features (e.g. the statistical properties of the spectral centroid for a sound being played back in forward or reverse are similar).
%
%
%
%
%
%
%
%
%
%
%
%
%
%
%
%
%
%
%\vspace{-0.2cm}
\section{Mapping \& Visualization}\label{mv}
%\vspace{-0.05cm}
%
%
%
%
A reduction of the high dimensional timbre space to two dimensions is deemed especially helpful in the context of multi-touch applications
since the lower dimensional representation can be directly used for visualization and interaction. 
As pointed out earlier, the disadvantage of using SOMs for this task is that they do not guarantee
certain properties for our use case, namely, the convergence of the algorithm and the preservation of topology. Therefore
we preferred Generative Topographic Mapping (GTM) \cite{Bishop1996}, exhibiting these characteristics. 
As a probabilistic method, the GTM defines a probability density function modeling how well a set of low L-dimensional latent variables, $x_k$, 
is mapped to high D-dimensional data-points $t_i$ by the function $y$. In our case, L is the 2D visualization space and D is the feature space. 
The mapping $y$ defines a non-linear transformation carried out by weighting Gaussian basis functions.
The centers of the these basis functions form a uniform grid in the latent space. So far, the intrinsic dimensionality of 
the mapping in data space is L. Only using this strictly confined L-dimensional manifold does not allow for some variance between the 
observed variables (the feature data) and the images of the latent variables. Therefore, the manifold is convolved with 
an isotropic Gaussian noise distribution with an inverse variance $\beta$ giving it some volume. These probabilistic properties allow to evaluate the 
quality of the mapping, such that the parameters $\beta$ and the weightings can be determined by Expectation Maximization, thus, generating the projection.
The continuous and smooth nature of $y$ allows for the topology preserving nature of the GTM - neighbor points in the latent 
space remain neighbor points in data space. This smoothness can be controlled by the parameters of the Gaussian basis functions set beforehand. 
At the end of the algorithm, the responsibility (probability) of each latent variable having generated each data point can be
evaluated. As latent variables are arranged on a grid, the position of each down-projected data point can be determined 
by weighting the latent variables' grid positions with their responsibility for it. In our case it is further necessary to perform a Z-Score transformation
of the feature data since the variance of the features diverge vastly (by the factor $\approx 10^9$) but the global parameter $\beta$ applies
to all dimensions. 

To help users differentiate more easily between the clusters of points in the projection, we performed a rough color coding to indicate cluster membership.
Since the GTM is topology preserving, one can apply K-Means Clustering in the feature space ($K\approx 50$) and assign colors accordingly.

Since we know which data point is projected onto which 2D position, we can construct an element wise bijective mapping.
For interaction, in order to find the nearest projected point in latent space from an arbitrary point (e.g.\ a user's touch) in the 2D space in real time, K-Nearest Neighbor search with $K=1$ using a KD-tree was applied. The resulting time-complexity of $\mathcal O(k \log n)$ with $\approx 10^6$ points permits comparatively quick look ups. 
In section \ref{dc} we will give an example of how we can benefit from more neighbors.
Finally, we applied a generic hash map to associate a feature vector with the respective set of generating parameters.
We will call the visualization, $p_i \rightarrow s_i \rightarrow t_i \rightarrow x_i$, and the mapping, $x_i \rightarrow t_i \rightarrow s_i \rightarrow p_i$, \textit{Timbre Surface} for future reference.

\begin{figure*}[htbp]
	\centering
		\includegraphics[width=\textwidth]{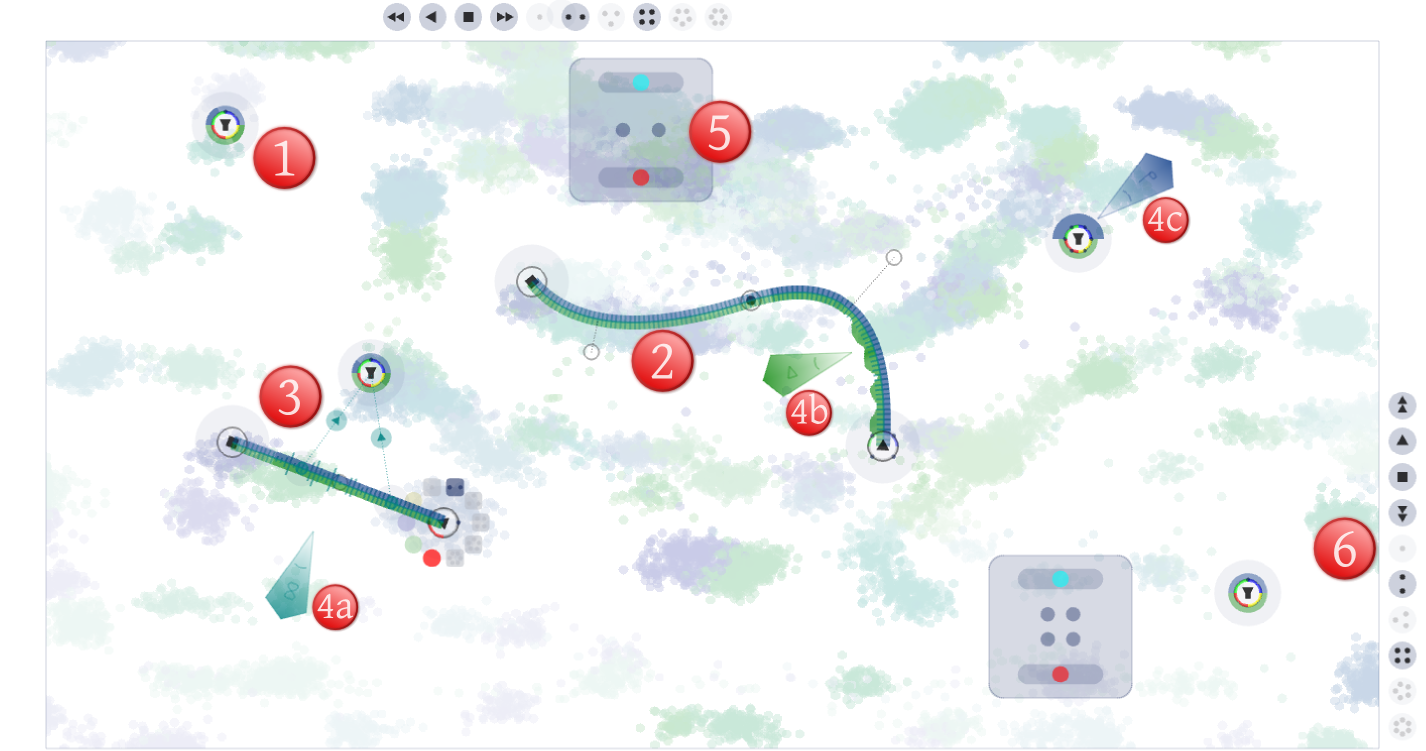}
	\caption{Screenshot of our prototype showing Nodes (1), Paths (2-3) and tools (4a-4c) as well as global controls for tonal parameters (5) and playback (6)}
	\label{outline}
%	\vspace{-0.4cm}
\end{figure*}
%\vspace{-0.2cm}
\section{Prototype}
%\vspace{-0.05cm}
We formulated the following constraints and requirements to frame the conceptual design of our prototypical application:\\
\textbf{Number of users:} The physical dimensions of our multi-touch table is 56", thus we limit the number of simultaneously collaborating users to five.\\
\textbf{Expertise of users:} As argued in the introductory section, the focus is to include novices but also experts into the collaborative sound design process. \\
\textbf{User objective:} We'd like to provide a tool for the explorative engagement with sounds but also their purposeful construction.\\
\textbf{Situational context:} Rather than to casually engage with the application, we expect users to get together purposefully. So we deem some explanatory exposition with respect to the interaction language and expressive possibilities permissible.\\
\textbf{Mode of Collaboration:} Possible modes among others are distributed leadership, turn-taking or a fully democratic process \cite{Blaine2003a}. We opted for the latter since this allows us to evaluate whether the proposed synthesis and interaction method aid users in comparison to a more conservative approach.\\
\textbf{Facilitation of Collaboration:} Besides using methodologies that aim to foster awareness, shareability and articulation, we aim at supporting the concept of public and private (acoustic) spaces and different levels of task coupling. Private spaces have been shown to give users more creative freedom to formulate contributions \cite{fencott2013computer}.%
\\
As stated earlier, users can design their sounds in a tonal context. For this, streams of harmonically fitting note events are generated and can be set to transpose sounds
within the limit of one octave. This way, we aim to preserve the validity of the timbral mapping \cite{Marozeau2003}
while providing a useful tonal extension.
Private acoustic spaces were supported by having
separate headphone output channels for every user that can be routed freely to hear any sound being played. Thus users are able to tune into the sound that collaborators design or set their sound public for the others to hear.

The Timbre Surface is at the core of the operational design. It spans the whole background of the user interface. Furthermore, we use the concept of a \textit{Node} (fig. \ref{outline}, item 1), a visually represented, draggable entity to facilitate the selection and playback of sounds according to its 2D position on the Timbre Surface. Nodes can be created and removed by simple gestures as seen fit and manipulated to change the sound's pitch and volume but also to set the incoming note stream and headphone output. These properties are directly
visually indicated to support awareness. 

Dragging a Node updates the synthesizer's parameters according to its position; the respective sound will be immediately heard. Continuous playback can be toggled with an additional button. 

To create sounds that change over time, Nodes can be connected to form Paths (fig. \ref{outline}, item 2). 
This expresses a timed motion over the Timbre Surface whose slope in between two Nodes can be altered with control points of a Bezier curve. This is the Timbre Space equivalent of parameter automation common in digital audio workstations. However, every Path represents a separate time-line. The segments of a Path are used for visualization, showing
the progress of playback but also for representation of the properties duration, volume and pitch at a certain point in time. 
By means of additional playback nodes and the possibility to chain paths (fig. \ref{outline}, item 3) more complex sequences of sounds can be created.

To modify the volume, pitch and time of a Path or Node, we followed a tool-based approach. This circumvents issues with ambiguous input (multi-touch gestures and parallel input) and
complex menu-driven command chains. For each of these three properties tools can be instantiated which can then be dragged and ``docked'' to the item to be modified. Then, by subsequent dragging of the separately highlighted tip of the tool, the property is modified accordingly. Figure \ref{outline} 4a, 4b, 4c show the tools for changing the duration of a segment, the volume and the pitch. 

A global menu can be used change the playback of all Nodes and Paths but also to show the widgets that manipulate the note streams. They show two parameters that modify e.g.\ the 
note length and frequency of an arpeggiated sequence of pre-programmed chord progressions. 

\begin{figure}\centering \includegraphics[width=1.0\columnwidth]{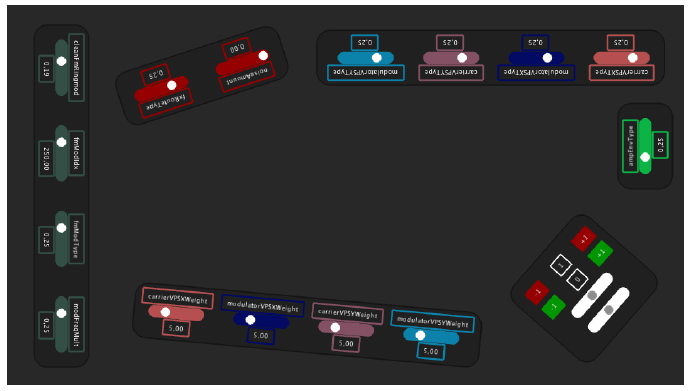}
	\caption{Screenshot of the alternative prototype}	

	\label{classic}
\end{figure}

\begin{figure}
\centering
\begin{minipage}{0.48\columnwidth}
\centering
\includegraphics[width=1\columnwidth]{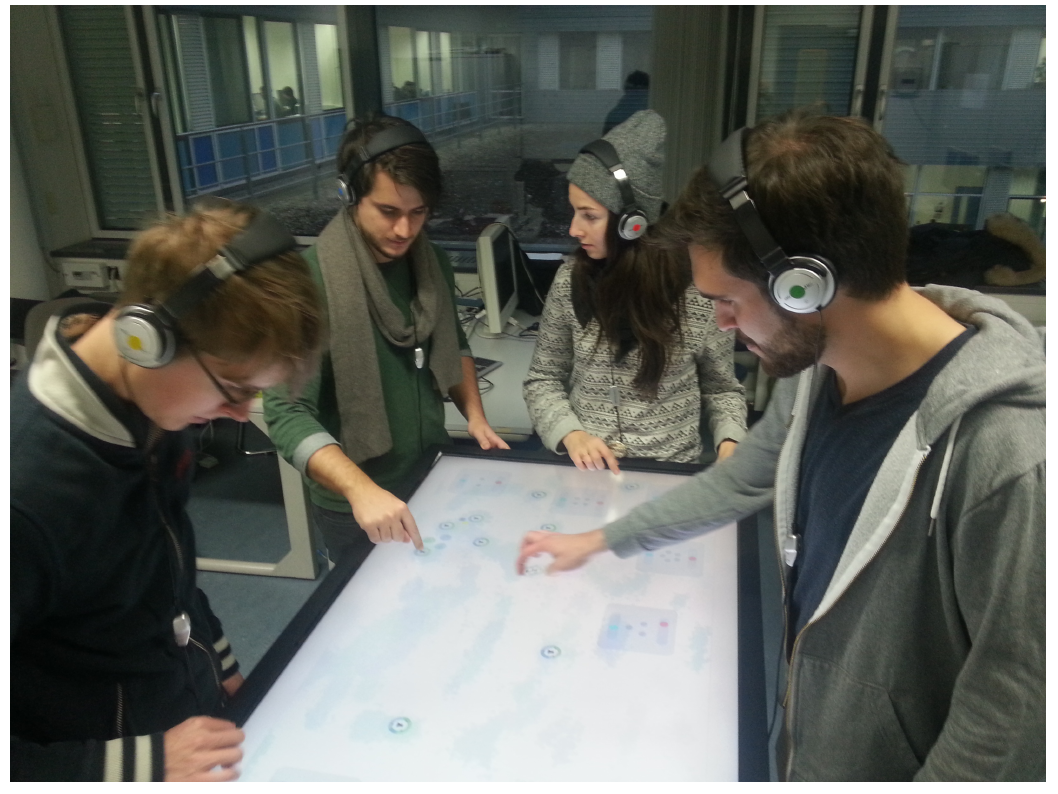}
\end{minipage}\hfill
\begin{minipage}{0.48\columnwidth}
\centering
\includegraphics[width=1\columnwidth]{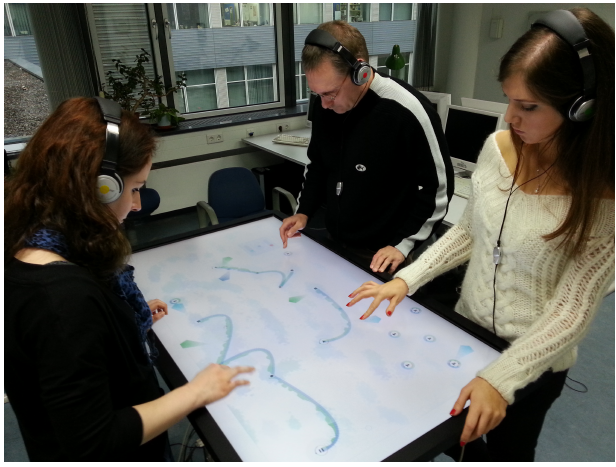}
\end{minipage}
\label{users}
\caption{Participants evaluating our prototype}

\end{figure}

\section{Evaluation}

We created a dataset containing 60,000 sounds, which %
is based on roughly 1,500 presets chosen by an expert and generated by the method described above. For the evaluation, we conducted a user study lasting one week that pursued the goal of comparing our approach to a classical one and to evaluate the practical design of the application with respect to user experience and collaboration. 

\subsection{Organization of the study}

The study was subdivided into three consecutive parts: comparison, experimentation
and questionnaire. 22 people took part it in groups of 2-4 collaborators. With respect to their demography, their age was between 19 and 43 years (mean 25.5) while 27\%
were female and 73\% male. Apart from one person, all participants enjoyed listening to music, most of them frequently, with a preference for Electronic and Rock/Pop. 
59\% of them had used music software before and 55\% played an instrument - 91\% also knew what a synthesizer was although only 27\% had used one before. 
14\% had composed a musical piece and 5\% designed sound. Regarding the technical knowledge, 59\% had previously used a multi-touch table. 
Only a small fraction of participants had prior ideas of what to accomplish creatively as the vast majority did not indicate any expectations towards the application or (collaborative) sound design.

For the comparative part, we developed an application that models a classic approach to sound design (cp fig. \ref{classic}) in a collaborative setting. 
We use the metaphor of a shared instrument where users can change only technical parameters of our synthesis method using simple sliders. These are grouped by functionality
as widgets that can be moved and rotated freely.

After a short explanation,
users were given 5 minutes to experiment with the classic approach and after that, again after an introduction, for the same duration to experiment
with the proposed one using only equivalent activities (node creation and dragging). We are well aware of the different approaches
to collaboration that both applications offer, however with respect to the functionality, we regard this comparison as permissible.
For the next part, we presented all of the remaining functionality and gave participants 25 minutes to delve deeper into the application but also 
to get accustomed to each other in hope that they would gradually focus on the collaborative aspects of sound-design (\ref{users}). A computer-mediated questionnaire
concluded the study. Alongside questions regarding demographic data, we addressed the comparison-, usability-, 
interaction- \& information design, visual design and finally the collaboration itself. Most questions used a 5-level Likert scale.

\subsection{Results}

The questionnaire showed mixed results regarding the overall ease of use, mental effort and 
the perceived ability to \textit{purposefully} execute ideas for our new application in comparison to the classical one. Corroborating the latter, participants 
were not always able to realise their musical and timbral ideas to a large part because they felt that finding specific sounds quickly was rather hard.
With respect to collaboration, it was stated that sharing and continuing work of others was not sufficiently \textit{facilitated} while 
it was \textit{perceived as easy}. This means that articulation was perceived to be fostered, as opposed to shareability. Reasons given for this were
first that wearing headphones was seen as a complication for communication, thus being detrimental to the process of collaboration. 
Yet more specific questions regarding the perception of related issues showed no significant statistical trend. 
Second, 
the time given for the experiment was not sufficient, as participants were still concerned with getting to know the application. Third, not knowing other participants 
led to less communication and interaction. 
Apart from these mixed results, the feedback for the application was very positive. The comparison between the classic and our approach showed that
ours was deemed more musically inspiring and incited people more to experiment and collaborate. It was further stated that it helped to obtain more musical results
both alone and in the group. The relationship between input and auditive output was found to be more understandable.
Somewhat surprisingly, but in line with the previous discussion, people perceived the new application as being able to provide more freedom in creating timbres although technically this is not the case. 
Regarding utility, most participants were able to find interesting timbres and create interesting complex ones.
Most interestingly, a minority stated that they had been shown new vistas with respect to music and timbre. 
In relation to engagement and pleasure, the
majority experienced Flow stating that they felt immersed. Collaboratively designing and experimenting 
with sound were rated as providing fun while it was felt that the collaboration in general had been fostered with enough access points to join in.
In terms of awareness and shareability, the possibility to experiment in a private auditory space was received very positively.
Finally, the application was favorably reviewed concerning the interaction and information design, as well as concerning aesthetics.

\section{Discussion \& Conclusion}\label{dc}

The evaluation revealed shortcomings with respect to:\\
\textbf{Navigation:} participants experienced a lack of insight how timbres
were arranged\\
\textbf{Awareness:} participants were confused as to who was doing what and which element
currently contributed to the overall sound.\\
Regarding the first shortcoming, users stated that they found Timbre Surface incoherent as small changes in position did not translate to small changes 
in timbre and that the clusters in the projection did not always have a comprehensible inter-relationship. The quality of the 
Timbre Surface depends crucially on the quality of the features. In this way the issue can be remedied with a different set of timbral features or 
different encodings thereof, such as the correlation of features for a sound. Additionally, the negative evaluation results with respect to the precision of the Timbre Surface led to a subsequent experiment investigating the influence of more thorough feature selection methods via observing
the likelihoods involved in the GTM. The results showed that these methods can improve the quality of the projection significantly. 
The GTM method itself also provides parameters that can be further adjusted. Real-time interpolation
of the sample points could lead to a smaller data-set and therefore disentangle the visualization but can also lead to a homogenization of parameter settings depending
on their neighborhood. For a quick evaluation of this method we used the nearest 8 neighbors of a position and weighted their parameters according to the Euclidean
distance from that position. This created a sufficiently smooth mapping where gradual changes in the 2D position led to gradual changes in the
aural output. However, this removed many of the original timbres from the Timbre Surface since this linear interpolation does not
inherit the non-linear nature of the GTM. Hence, a more complex interpolation method is needed which takes the GTM into account. For example,
the gradient of the generated responsibilities could be used for weighting. Another approach is to introduce complementary UI tools, that can pull clusters apart (e.g.\ a ``magnifing glass''). \\
With respect to the second shortcoming, participants stated that the issue is mainly the control over the private space and the related default settings used.
As suggested by many participants themselves, this issue can be resolved by providing users the option to mute specific co-users altogether or to provide a user-specific
mixer functionality. Removing private spaces, however, does not appear to be an option as this was regarded a \textit{necessity} by a large majority
of the participants.\\
To summarize the positive feedback: \\
\textbf{Comparison to classic approach:} Our application was received as a valuable alternative to the classic approach to sound-design, being perceived as
more musical, expressive, inspiring, comprehensible and inciting towards collaboration.\\
\textbf{Utility and Pleasure:} Participants stated that they were able to find interesting and design complex timbres. This activity was perceived as
pleasurable and Flow inducing. Furthermore, the collaborative use-case is seen as fun.\\
\textbf{Collaboration:} The application as been reviewed as supporting and fostering collaboration.\\
Although the named shortcomings conflict with the goals set at the beginning of this paper, we do not assess them as overly severe since they are not conceptual issues but 
rather technical ones that can be approached systematically within further research. 
To conclude, given this positive feedback and the amount of committed suggestions for improvements by participants, we see the results as satisfactory.
In this regard, we furthermore would like to point out that this application has been developed with our previously conducted research in mind,
namely the automatic emotional affect estimation of timbres \cite{Klugel2013}. Here, sounds can be given an affect value (Valence and Arousal) that can be included in the %
feature vector for projection. 
Thus allowing to design sound not only with timbral descriptions in mind but also with emotional affect related ones.
In spirit of our previous work, the data-sets (feature- \& parameter data, evaluation) and source-code of our application and frameworks are available from our website\footnote{\url{https://github.com/lodsb/UltraCom}}.
\bibliographystyle{abbrv}

\bibliography{lib2.bib}  

\begin{thebibliography}{10}

\bibitem{Bishop1996}
C.~Bishop~et al.
\newblock {G}{T}{M}: A principled alternative to the self-organizing map.
\newblock In {\em Proc. Int Conf. Artif. Neur. Net.}, pages 165--170. Springer,
  1996.

\bibitem{Blaine2003a}
T.~Blaine~et al.
\newblock {Collaborative Musical Experiences for Novices}.
\newblock {\em J. New Music Res.}, 32(4), Dec. 2003.

\bibitem{Cockburn:1995}
A.~Cockburn~et al.
\newblock {Four principles of groupware design}.
\newblock {\em Interact. Comput.}, 7(2), 1995.

\bibitem{Comajuncosas2011}
J.~M. Comajuncosas~et al.
\newblock {Nuvolet : 3D Gesture-driven Collaborative Audio Mosaicing}.
\newblock {\em Proc. Int. Conf. NIME}, 2011.

\bibitem{Cottrell2004}
S.~Cottrell.
\newblock {\em {Professional music-making in London : ethnography and
  experience}}.
\newblock Ashgate, 2004.

\bibitem{Csikszentmihalyi1990}
M.~Csikszentmihalyi.
\newblock {\em {Flow: The Psychology of Optimal Experience}}.
\newblock Harper and Row, 1990.

\bibitem{Dourish:1992}
P.~Dourish~et al.
\newblock {Awareness and coordination in shared workspaces}.
\newblock In {\em Proc. Int. Conf. CSCW}, 1992.

\bibitem{Eigenfeldt}
A.~Eigenfeldt~et al.
\newblock Realtime timbral organisation: Selecting samples based upon
  similarity.
\newblock {\em Organised Sound}, 15(2), 2010.

\bibitem{fencott2013computer}
R.~Fencott~et al.
\newblock Computer musicking: Hci, cscw and collaborative digital musical
  interaction.
\newblock In {\em Music and Human-Computer Interaction}. Springer, 2013.

\bibitem{Fischer2005}
G.~Fischer.
\newblock {Distances and diversity: sources for social creativity}.
\newblock In {\em Proc. 5th Conf. Creat. Cogn.}, pages 128--136. ACM, 2005.

\bibitem{Fischer2006}
G.~Fischer.
\newblock Distributed intelligence: extending the power of the unaided,
  individual human mind.
\newblock In {\em Proc. Int. Conf. Adv. Vis. Interf.} ACM, 2006.

\bibitem{Hoffman}
M.~Hoffman~et al.
\newblock Feature-based synthesis: Mapping acoustic and perceptual features
  onto synthesis parameters.
\newblock In {\em Proc. Int. Conf. ICMC}. Citeseer, 2006.

\bibitem{Holmes2011}
P.~Holmes.
\newblock {An exploration of musical communication through expressive use of
  timbre: The performer's perspective}.
\newblock {\em Psychol. Music}, Mar. 2011.

\bibitem{Hornecker2007}
E.~Hornecker~et al.
\newblock From entry to access: how shareability comes about.
\newblock In {\em Proc.~Conf.~Des. Pleasurable Prod. Interf.} ACM, 2007.

\bibitem{Kleimola2011}
J.~Kleimola~et al.
\newblock {Vector Phase Shaping Synthesis}.
\newblock {\em Proc. Int. Conf. DAFX}, 2011.

\bibitem{Klugel2013}
N.~Kl\"{u}gel~et al.
\newblock {Towards Mapping Timbre to Emotional Affect}.
\newblock {\em Proc. Int. Conf. NIME}, 2013.

\bibitem{Lartillot2008}
O.~Lartillot~et al.
\newblock A matlab toolbox for music information retrieval.
\newblock In {\em Data analysis, machine learning and applications}. Springer,
  2008.

\bibitem{LeGroux2008}
S.~{Le Groux}~et al.
\newblock {Perceptsynth: mapping perceptual musical features to sound synthesis
  parameters}.
\newblock {\em Proc. Int. Conf. Acoust. Speech Sig. Process.}, 2008.

\bibitem{MacDonald2006}
R.~MacDonald.
\newblock {Creativity and flow in musical composition: an empirical
  investigation}.
\newblock {\em Psychol. Music}, 34(3), July 2006.

\bibitem{Marozeau2003}
J.~Marozeau~et. al.
\newblock The dependency of timbre on fundamental frequency.
\newblock {\em J. of Acoust. Soc. America}, 114(5), 2003.

\bibitem{Morgan1980}
G.~Morgan.
\newblock {Paradigms, Metaphors, and Puzzle Solving in Organization Theory}.
\newblock {\em Adm. Sci. Q.}, 25(4), 1980.

\bibitem{Nicol2004}
C.~Nicol~et al.
\newblock Designing sound: Towards a system for designing audio interfaces
  using timbre spaces.
\newblock In {\em ICAD}, 2004.

\bibitem{Nijs2009}
L.~Nijs.
\newblock {\em {The musical instrument of natural extension of the musician}}.
\newblock Witwatersrand Univ. Press, 2009.

\bibitem{Pampalk2004}
E.~Pampalk.
\newblock A matlab toolbox to compute music similarity from audio.
\newblock In {\em Proc. Int. Conf. ISMIR}, 2004.

\bibitem{Risset2003}
J.-C. Risset.
\newblock The perception of musical sound.
\newblock 2003.

\bibitem{Robel}
A.~R\"{o}bel.
\newblock Between physics and perception: Signal models for high level audio
  processing.
\newblock {\em Proc. Int. Conf. DAFX}, 2010.

\bibitem{Sawyer2006}
R.~K. Sawyer.
\newblock {Group creativity: musical performance and collaboration}.
\newblock {\em Psychol. Music}, 34(2):148--165, Apr. 2006.

\bibitem{Schwarz2012}
D.~Schwarz.
\newblock The sound space as musical instrument: Playing corpus-based
  concatenative synthesis.
\newblock {\em Proc. Int. Conf. NIME}, 2012.

\bibitem{seago2013new}
A.~Seago.
\newblock A new interaction strategy for musical timbre design.
\newblock In {\em Music and human-computer interaction}. Springer, 2013.

\bibitem{Serra1997}
X.~Serra.
\newblock {Musical sound modeling with sinusoids plus noise}.
\newblock {\em Music. signal Process.}, pages 1--25, 1997.

\bibitem{Sternberg1999}
R.~J. Sternberg.
\newblock {\em {Handbook of creativity}}.
\newblock Perspectives on individual differences. Cambridge University Press,
  1999.

\bibitem{Swift2004}
B.~Swift~et al.
\newblock {Engagement Networks in Social Music-making}.
\newblock {\em Proc. Int. Comp. Hum. Interact.}, 2004.

\bibitem{Uzzi2005}
B.~Uzzi~et al.
\newblock {Collaboration and Creativity: The Small World Problem}.
\newblock {\em Am. J. Sociol.}, 111(2), Sept. 2005.

\bibitem{Wigdor2011}
D.~Wigdor~et al.
\newblock {\em {Brave NUI World}}.
\newblock Morgan Kaufmann, 2011.

\bibitem{Zolzer2003}
U.~Z\"{o}lzer and J.~{Smith Iii}.
\newblock {DAFX: Digital Audio Effects}.
\newblock {\em J. Acoust. Soc. Am.}, 114:2527, 2003.

\end{thebibliography}
\end{document}